# Anisotropic Electron-Phonon Coupling and Dynamical Nesting on the Graphene Sheets in $CaC_6$

T. Valla[1*], J. Camacho[1], Z-H. Pan[2], A. V. Fedorov[2], A. C. Walters[3], C. A. Howard[3], M. Ellerby[3]

[1] *Condensed Matter Physics and Materials Science Department, Brookhaven National Laboratory, Upton, NY 11973, USA.*

[2] *Advanced Light Source, Lawrence Berkeley National Laboratory, Berkeley, CA 94720, USA.*

[3] *London Centre for Nanotechnology and Department of Physics and Astronomy, University College London, London WC1E 6BT, United Kingdom*

*\* e-mail:valla@bnl.gov*

Superconductivity in graphite intercalated compounds (GIC) has been studied for more than 40 years[1,2] and it is still not fully understood, despite the recent progress and the discovery[3,4] of relatively high Tc superconductivity in $CaC_6$ and $YbC_6$. Recent studies now suggest that the electron-phonon coupling (EPC) is most likely responsible for pairing[5,6,7] and that the intercalant-derived electronic states and vibrations play the dominant role.[8,9,10,11,12] Here, we present the first studies of electronic structure in $CaC_6$, a superconductor with Tc=11.6 K. Using angle-resolved photoemission spectroscopy (ARPES), we find that, contrary to theoretical models, the EPC on the graphene-derived Fermi sheets is surprisingly strong, reflecting the interaction with high-frequency graphene-derived vibrations. Thus, in addition to the amazing properties in the charge-neutral state,[13] graphene sheets also show surprises in the heavily doped regime: they may support strong

**pairing interactions and lead to superconductivity in compounds in which they are building blocks.**

In conventional metals, EPC has long been known to be the pairing interaction responsible for the superconductivity. The strength of this interaction essentially determines the superconducting transition temperature Tc. One manifestation of EPC is a renormalization of the electronic dispersion or a "kink" at the energy scale associated with the phonons. This renormalization is directly observable in ARPES experiments.[14] In GICs, the EPC is now considered to be the pairing mechanism that leads to relatively high Tc. Most of the theoretical studies suggest that the intercalant-derived Fermi surface (FS)[5] and intercalant-derived soft vibrations[9,10] are the key for superconductivity, while some suggest that the intercalant states couple also to the out-of-plane graphene vibrations.[11,8] However, all of them agree that the graphene electrons are only very weakly coupled to graphene vibrations, rendering this interaction virtually irrelevant for superconductivity. Here, we show that this interaction is very strong and anisotropic and that it alone might be sufficient to induce superconductivity in GICs. We give a possible explanation for this enhancement and discuss its role in superconductivity.

Fig 1. shows the ARPES spectra near the K point in the graphene Brillouin zone (BZ). The upper panel (a) shows the contours of photoemission intensity as a function of binding energy for several momentum lines in the vicinity of K point, while the intensity from a narrow interval (±10 meV) around the Fermi level is shown in the lower panels (b and c). For comparison, we also show the ARPES spectrum from the momentum line going through K point (d) and the spectral contours at the Fermi level (e) and at $\omega=-1eV$ (f) of a thin (~7-9 graphene layers), highly oriented pyrolitic graphite (HOPG) flake on a $SiO_2$ substrate. While the FS of the graphite flake collapses nearly into a point - Dirac point - the FS of $CaC_6$ encloses a significant area and has a concave triangular shape centred at the K point of the BZ. The charge transfer from intercalant atoms fills the graphene bands and the Dirac point is moved to ~1.5 eV below the Fermi level. Two important observations about $CaC_6$ can be made: 1) in all the spectra from Fig. 1a, there is a renormalization of the quasiparticle dispersion at ~160 meV below the Fermi level

and 2) the contour of the spectral intensity at the Fermi level (Fig. 1b and 1c), corresponding to the grapheme-derived FS, is smaller than what would be expected if Ca were completely ionized. The area enclosed by these contours corresponds to the doping level of graphene sheets and in the case of complete Ca ionization, this should be 1/3 of the area of the BZ, equivalent of the doping of 1/3 electrons per C atom. Instead, the measured Fermi area gives 0.09 electrons per C atom. Also, the measured binding energies of Ca 3s and 3p core levels (43.76 eV and 24.63 eV, respectively) are much closer to those in the metallic calcium than in the model ionic compounds such as CaO or $CaF_2$, indicating that the charge transfer from Ca is not complete.[15,16]

It is apparent from Fig. 1 that the renormalization effects ("kinks") are not equally strong for all the points on the FS of CaC6 and that they are much stronger than in graphite. Fig. 2 illustrates the anisotropy of the renormalization effects in $CaC_6$. We have extracted the dispersions and the linewidths of the quasiparticle states by fitting the spectral intensities at constant energies, or momentum distribution curves (MDCs),[14] with Lorentzian distributions. The real and imaginary components of the self-energy are then derived in the usual manner from the peak positions, $k_m$, and widths, $\Delta k$, of the MDC peaks using the expressions

$$\mathrm{Re}\Sigma = (k_m - k)v_0, \qquad \mathrm{Im}\Sigma = \frac{\Delta k}{2}v_0 \qquad (1)$$

We have approximated the bare dispersions with straight lines, $\varepsilon_k = v_0(k-k_F)$ where $v_0$ represents the non-interacting Fermi velocity for a given line in momentum space. The results for $\mathrm{Re}\Sigma$ and $\mathrm{Im}\Sigma$, corresponding to the polar angle $\varphi=17°$ and $\mathrm{Im}\Sigma$ for $\varphi=25°$ are shown in Fig. 2c). Fig. 2d) shows $\mathrm{Re}\Sigma$ for different momentum lines corresponding, from bottom to top, to polar angles $\varphi=7°, 28°, 45°$, and $58°$. Self-energies have a structure typical for the interaction with a well defined bosonic mode: the imaginary part has a sharp step-like feature, while the real part has a peak at the energy of the mode, $\Omega_0$. Therefore, the mode energy can be read from the measured self-energy. In our case, the dominant structure occurs at ~160 meV, with an additional feature at ~75 meV. These features can be naturally attributed to the interaction with in-plane and out-of-plane phonons of the graphene sheets.

We note that the finite energy resolution affects the measured ReΣ near the Fermi level, within the resolution range.[17] We have therefore excluded the affected interval |ω|<20 meV from the considerations and any fine structure, related to a possible coupling to the intercalant modes, is out of our detection limits. At higher energies, we can model the measured self-energy at any φ with an Eliashberg function, $\alpha^2 F(\omega)$, consisting of two peaks: one fixed at ω=75 meV and the other ranging from 155 to 165 meV. The relative contribution of these two peaks increases from 1:5 to almost 1:1 on moving from φ≈0 to φ≈60°. The modelled ReΣ are shown for these two limiting cases. Our model assumes a constant density of states (DOS), whereas a linear DOS would be a better approximation since the scattering is expected to reflect the density of final states. This is visible in Fig. 2c, where ImΣ quickly acquires a slope as one moves from ΓK to KM. Near the KM lines, the high-energy side of ReΣ also deviates from the simple model. However, in the energy range 20≤|ω|≤ 200meV, the model works well and the fitting of measured ReΣ gives the energy of the high frequency mode $\Omega_0$. In addition, the coupling strength, λ, can be extracted from the low energy slope of ReΣ, $\lambda = -(\partial \text{Re}\Sigma/\partial \omega)_{\omega \to 0}$. From Fig. 2d) it is clear that the characteristic energy $\Omega_0$ does not vary a lot, whereas the coupling strength λ displays a very strong anisotropy. In Fig. 3, we plot both quantities as functions of polar angle φ. The coupling constant ranges from 0.39 (for φ≈0) to 1.54 (for φ≈60°), with the momentum averaged value of 0.67. The coupling constant in pristine graphite flake is much smaller (λ=0.18) and does not show a significant anisotropy, in agreement with Leem *et al.*[18] Recent theoretical calculations suggest that in $CaC_6$, the crucial role is played by calcium, i.e. electrons on the intercalant-derived FS are strongly coupled to the intercalant vibrations.[9,10,11,8] Those calculations show ~one order of magnitude weaker coupling of the graphene-derived electrons to the in-plane graphene phonons, with essentially no anisotropy.[19] The large calcium isotope effect also implies the dominant role of soft Ca vibrations in superconductivity.[7] Our results, on the contrary, suggest that graphene sheets play the dominant role because the obtained EPC on graphene-derived FS with the in-plane and out-of–plane graphene phonons seems to be more than sufficient to give Tc≈11.6 K.

Where is such strong and anisotropic coupling coming from? Similar anisotropy is recently reported for EPC in K and Ca doped graphene layers on SiC.[20] There, the proximity of van Hove singularity (vHS) to the Fermi level is suggested to be an origin of the enhanced EPC. While this may certainly play a role, especially for the small momentum or *intra-valley* scattering, we have observed that in the case of $CaC_6$, very favourable dynamical nesting conditions exist that may enhance the *inter-valley* EPC. This is illustrated in Fig. 3c). In order to satisfy the energy conservation rule in an electron-phonon scattering event, the photohole created at ~160 meV has to be scattered to the Fermi level after emission of a 160 meV phonon. To explore what momenta are allowed in such a process, we plot the states at $\omega=-160$ meV at one corner of the BZ and the states at the Fermi level in another corner. A wavevector connecting any pair of points on different contours represents an allowed scattering process. Note that in the case of *intra-valley* scattering, both contours would have to be centred at the same K point. However, from the measured phonon spectra in graphite,[21] and recent Raman experiments on $CaC_6$,[22] it is clear that there are no $q\sim0$ modes at ~160 meV and we have to consider *inter-valley* scattering. From Fig. 3c), it appears that some wave-vectors are particularly efficient in connecting the two contours. As the $\omega=-160$ meV contour is slightly convex and the FS is slightly concave with nearly the same curvature, the green vector in Fig 3c) effectively nests the whole one side, or ~1/3 of the length of the two contours. The tips of contours can be nested by two such vectors, which would explain a portion of the observed anisotropy in $\lambda$. The additional factor characterizing the tips is a gain in coupling to lower frequency phonons at $\omega\approx75$ meV, as can be seen in Fig. 2d. We note that the observed "nesting" is not static: it involves emission/absorption of a high frequency phonon and would not produce the charge density wave (CDW) instability. This dynamical nesting and resulting enhancement of EPC is likely to be very sensitive to band filling. For lower doping levels, the nesting efficiency might be lost as both contours become convex. At higher fillings, the coupling to lower frequency phonons near the KM lines could eventually drive the system into the CDW state.

The measured Fermi velocity on the graphene-derived FS is also very anisotropic: $v_F(\varphi=180°)\approx 1.1$ eVÅ, $v_F(\varphi=0)\approx 4.1$ eVÅ, with the momentum average $<v_F>\approx 2.5$ eVÅ. If the graphene electrons were responsible for superconducting properties, the in-plane coherence length, $\xi_{ab}=\hbar v_F/(\pi\Delta)$, would be in the range 200 Å $< \xi_{ab} <$ 800 Å, if we take $\Delta\approx 1.6$ meV for the superconducting gap.[23] $<v_F>$ gives $\xi_{ab}\approx 500$ Å, in reasonable agreement with $\xi_{ab}\approx 350$ Å reported in STM,[23] specific heat[5] and magnetization[4,6] studies. We also note that peculiar linearity in the upper critical filed, $H_{c2}$, at low temperatures, could also be explained by anisotropic Fermi velocity on the graphene-derived FS.[8] In contrast, the highly isotropic intercalant FS would not produce linear $H_{c2}$.

The remaining question is the absence of an intercalant-derived FS in our experiments. These states are predicted to be free-electron-like, with a 3-dimensional, nearly spherical FS. We have searched at several photon energies, ranging from 40 to 75 eV, but found no evidence of the intercalant-derived FS. However, our results show that, irrespectively of whether this band is occupied or not, the measured EPC on the graphene-derived FS to the high frequency graphene phonons alone might be more than sufficient to explain relatively high Tc in $CaC_6$.

## *Methods*

**Samples:** The $CaC_6$ samples were prepared via immersion of a graphite platelet in a lithium/calcium alloy for 10 days, as described in detail by Pruvost *et al*.[24] The graphite used was HOPG, or natural, single-crystal graphite flake (Madagascan). The lithium and calcium were purchased from Sigma-Aldrich at 99.99% purity. X-ray diffraction of the resulting shiny silver-like $CaC_6$ platelets showed very high sample purity with no graphite or $LiC_6$ impurity phases within the limits of our measurements (i.e <1%). SQUID magnetometry measurements on the samples revealed sharp ($\Delta T<0.3K$) superconducting transition at ~11.6 K (onset). To avoid degradation, samples were unsealed and glued to the sample holder with Ag-epoxy in an Ar filled glow box. Samples, protected by the cured epoxy, were then transferred to the ARPES prep-chamber. The exposure to air was limited to a fraction of a minute before the pump-down. In regard to possible sample-to-sample variations in ARPES experiments, we have obtained identical results from two different $CaC_6$ flakes and very similar renormalizations and dispersions on the intercalated HOPG sample in the explored range of the BZ.

The thin flake of pristine graphite was obtained by micromechanical exfoliation of HOPG on $SiO_2$/Si substrate in air. From the optical contrast and from Raman data, we determined the thickness of the flake to be in the range of 7-9 graphene layers. Based on the dispersion, we estimate that the Dirac point in the flake is at $\omega=10\pm10$ meV.

**Angle-Resolved Photoemission Spectroscopy** experiments were carried out on a Scienta SES-100 electron spectrometer operating in the angle-resolved mode at the beamline 12.0.1 of the Advanced Light Source. $CaC_6$ samples were mounted on a high precision variable-temperature goniometer with three angular and three linear degrees of freedom and cleaved at low temperature (~15-20 K) under ultra-high vacuum (UHV) conditions (~$1\times10^{-9}$ Pa). The pristine graphite flake was annealed to ~600 K in the UHV chamber before ARPES studies. The spectra presented here were recorded at the photon energy of 45 eV, with the combined instrumental energy resolution of ~ 20-25 meV and the momentum resolution of $\pm0.008$ Å$^{-1}$ in geometry where the polarization of light was perpendicular to the probed momentum line and analysed using momentum distribution curves (MDC). The temperature was measured using a calibrated silicon sensor mounted near the sample.

## *Acknowledgements*

We acknowledge technical help from Antony Bollinger and useful discussions with Phil Allen, Peter Johnson, Myron Strongin, Alexei Tsvelik and Mary Upton. The research work described in this paper was supported by the US Department of Energy and the UK Engineering and Physical Science Research Council.

Correspondence and requests for materials should be addressed to T.V.

## *Competing financial interests*
The authors declare that they have no competing financial interests.

*Figures:*

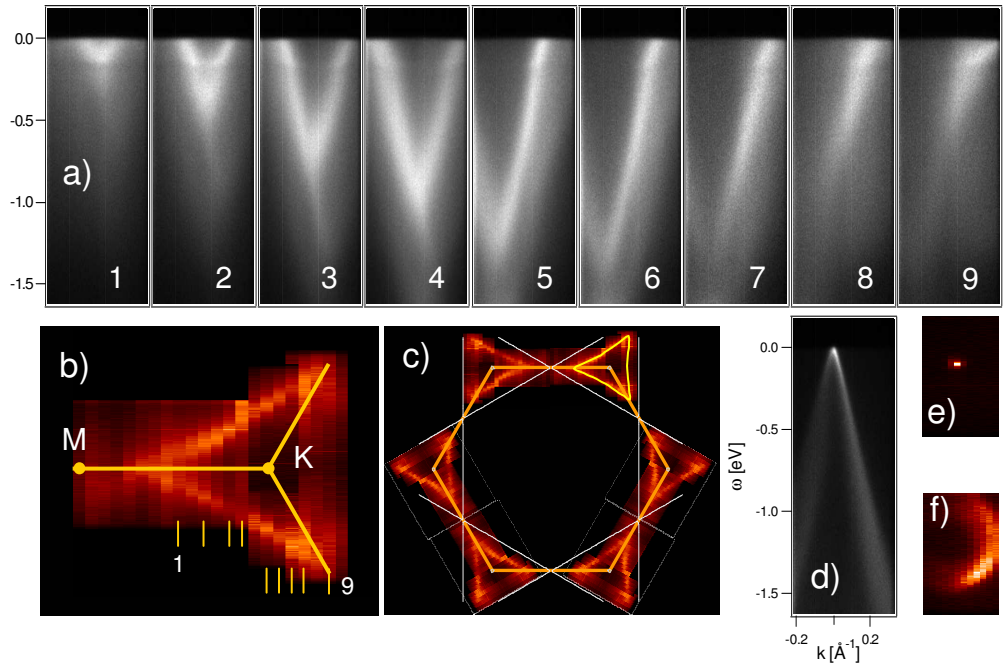

**Figure 1. ARPES from $CaC_6$ and thin graphite. a)** 1-9: photoemission spectra from $CaC_6$ flake from several momentum lines near the K point as indicated in b). **b)** Photoemission intensity from $CaC_6$ from a narrow energy interval around the Fermi level ($\omega=\pm 10$ meV) near the K point of the BZ. Thin lines indicate momentum lines probed in a) **c)** Fermi surface of $CaC_6$ obtained by 6-fold symmetryzation of intensity shown in b). White lines correspond to the doping of ¼ electrons per C atom, at which the van Hove singularity in the $\pi^*$ band sits at the Fermi level. **d)** ARPES spectrum from a thin graphite flake from the momentum line going through the K point orthogonal to the K-M line. **e)** photoemission intensity from the graphite flake around K point at $\omega=0$ and **f)** at $\omega=-1$eV. All the spectra in a), b), d), e) and f) are shown on the same momentum and energy scales and all were taken at T=14 K, in the normal state of $CaC_6$.

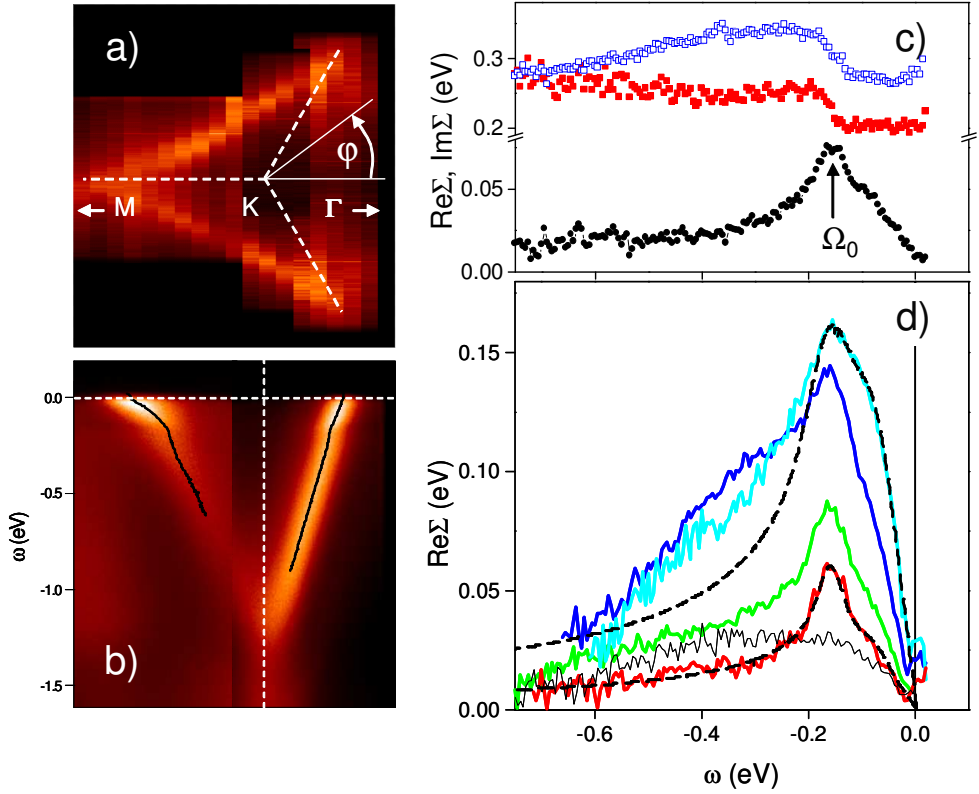

**Figure 2. Quasiparticle self-energy on the graphene-derived Fermi surface. a)** Fermi surface of $CaC_6$. The dashed lines correspond to KM lines in the BZ. $\varphi$ defines the polar angle around the K point, measured from the $\Gamma$K line. **b)** Spectral intensities for $\varphi=0$ or $\Gamma$K line (left) and $\varphi=180°$ or KM line (right). Black solid lines represent the MDC derived dispersions. c) Re$\Sigma$ (black circles) and Im$\Sigma$ (red squares) for $\varphi = 17°$ and Im$\Sigma$ for $\varphi = 25°$ (blue squares). Peak (step) position in Re$\Sigma$ (Im$\Sigma$) is labeled with $\Omega_0$. d) Real part of the self-energy for several points on the Fermi surface, $\varphi = 7°, 28°, 45°,$ and $58°$, (bottom to top). Dashed lines represent the fits to the simple model self-energies as described in text. Thin black line is Re$\Sigma$ of pristine graphite flake.

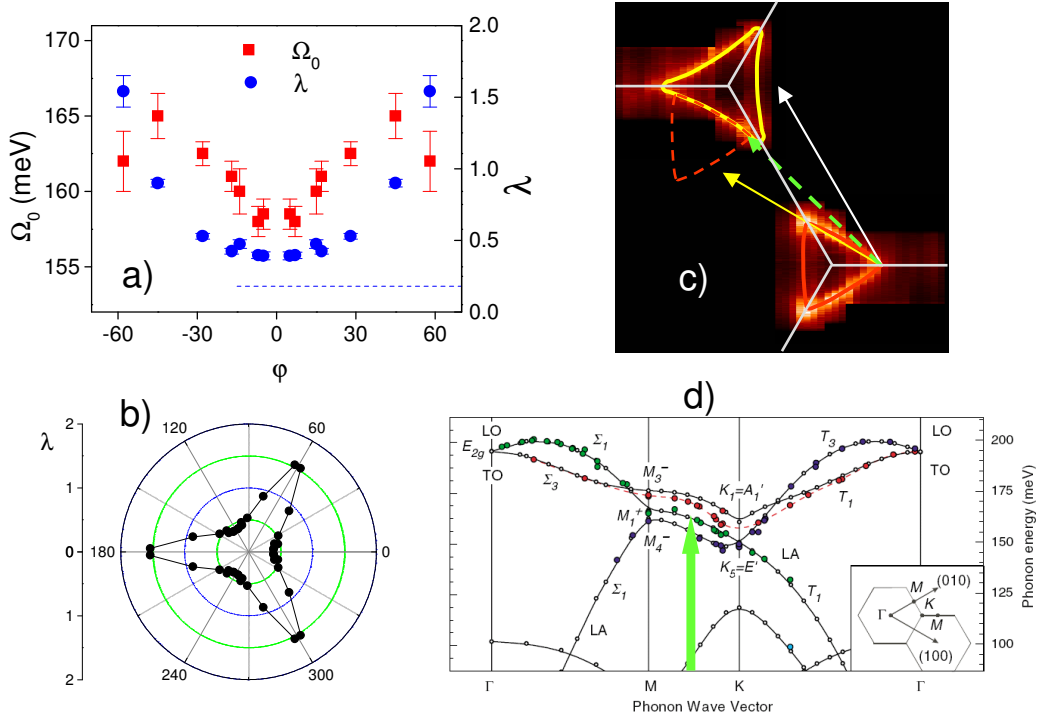

**Figure 3. Anisotropic electron-phonon coupling in CaC$_6$.** a) Characteristic energy $\Omega_0$ of the maximum in Re$\Sigma$ and the coupling strength $\lambda$, extracted from Re$\Sigma$ as functions of polar angle $\varphi$ as defined in Fig.2. The dashed blue line represents the EPC for pristine graphite flake. **b)** Polar plot of the EPC constant, $\lambda$. c) Dynamical nesting of the spectral contours at $\omega=-160$ meV (red contour) and $\omega=0$ (yellow contour). White (yellow) arrow represents the $\Gamma$K ($\Gamma$M) wavevector, while the dashed green arrow represents the wavevector that efficiently "nests" the two contours. d) Phonon dispersions in pristine graphite from ref. 21. The position of "nesting" vector from c) is marked with green arrow whose length corresponds to the exchanged energy.